\documentclass[11pt,twoside]{article}
\usepackage{asp2014}

\aspSuppressVolSlug
\resetcounters

\begin{document}

\title{Stability of an active longitude of the giant PZ~Mon}
\author{Pakhomov Yu.~V.$^1$, Antonyuk K.~A.$^2$, Bondar' N.~I.$^2$, Pit N.~V.$^2$
\affil{$^1$Institute of Astronomy, Russian Academy of Sciences, Moscow, Russia; \email{pakhomov@inasan.ru}}
\affil{$^2$Crimean Astrophysical Observatory RAS, Nauchny, Republic of Crimea, Russia; \email{otbn@mail.ru}}
}

\paperauthor{Pakhomov Yu.}{pakhomov@inasan.ru}{}{Institute of Astronomy, Russian Academy of Sciences}{}{Moscow}{}{117019}{Russia}
\paperauthor{Antonyuk K.A.}{antoniuk@craocrimea.ru}{}{Crimean Astrophysical Observatory RAS}{}{Nauchny}{Republic of Crimea}{298`409}{Russia}
\paperauthor{Bondar' N.I.}{otbn@mail.ru}{}{Crimean Astrophysical Observatory RAS}{}{Nauchny}{Republic of Crimea}{298409}{Russia}
\paperauthor{Pit N.V.}{petersola@mail.ru}{}{Crimean Astrophysical Observatory RAS}{}{Nauchny}{Republic of Crimea}{298409}{Russia}

\begin{abstract}
Analysis of photometric data of the active giant PZ~Mon is presented. Using ASAS-3 project data and new more accurate photometry we establish that during 15 years of PZ~Mon CCD observations the light curve remains stable, and consequently a longitude of the active spotted area is stable. The small deviations may be explained by differential rotation or inhomogeneous distribution of spots on the active hemisphere of PZ~Mon. The stability of the active longitude and it's location on the PZ~Mon surface indicates on the secondary component as reason of stellar activity. 
\end{abstract}

\section{Introduction}

PZ~Mon\ (HD\,289114, $V\approx9$~mag) is an active K2III star of RS~CVn type with the radius of 7.7~$R_\odot$ and the mass of 1.5~$M_\odot$ \citep{2015MNRAS.446...56P}. The secondary component is a cool dwarf with the radius of 0.15~$R_\odot$ and the mass of 0.14~$M_\odot$ which moves on the circular orbit with the major semiaxis of 0.24~au \citep{2015AstL...41..677P}. Fig.~\ref{model} shows the scaled model of the PZ~Mon system.

\articlefigure{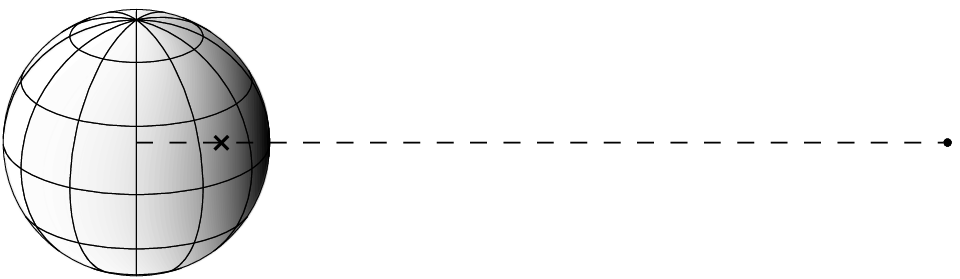}{model}{Scaled model of the PZ~Mon system. Centre of mass is marked by the cross. Darker areas on the surface of the main component corresponds to the higher density of spots.}

The photometric behaviour of the variable star PZ~Mon is not ordinary, it reflects the complex stellar activity. The global cycle is presented by the amplitude of about 1$^m$ and the period of $\sim$50~years \citep{1995AAS..111..259B}. Also there are small cycles with the amplitudes of about 0.1$^m$ and the periods $\sim$22 and $\sim$6.7~years \citep{2007OAP....20...14B}. These periods were found from measurements of the photographic plates taken since 1899 in several European and Russian observatories. The special photometric study was not planned, therefore, the smallest periods were not found. Further research is performed using CCD. The most complete modern photometry of PZ~Mon is contained in ASAS-3 project \citep{1997AcA....47..467P}, but it's accuracy (0.02 -- 0.05$^m$) is not enough to the detailed study of the stellar surface. Nevertheless, the photometric period of 34.13$\pm$0.02~days with the amplitude of about $0.01-0.05^m$ was determined. This period is most expressed and always present that allow us to attribute it for rotational modulation of the spotty star. The joint analysis of light curve and radial velocity curve led us to conclusion that spotted area on the PZ~Mon surface is located under the secondary component (Fig.~\ref{model}).

\articlefigure{asas.eps}{asas}{ASAS-3 photometry of PZ Mon. \textit{Top:} All data in years 2001--2009. \textit{Middle:} Data sets during 2007-2008 with a variable amplitude. \textit{Bottom:} Phase curve of data sets in 2005-2009 which is reduced by main trend and approximated by cosines law.}

Radial velocity period 34.15$\pm$0.02~days is very close to the photometric that may be explained as synchronisation of the rotation and orbital motion in the PZ~Mon system. The mass ratio of the components is 0.09$\pm$0.03 -- the smallest value among the known RS~CVn type giants \citep{2015AstL...41..677P}. The nature of the synchronisation by a component of small mass is unknown. So, it is important to check the fact of synchronisation and stability of the active longitude by new more accurate photometric data.  

\section{Observations and analysis of the position of active longitude}

We used photometric data of ASAS-3 project and new own observations. ASAS photometry data of PZ~Mon covers the dates in 2001--2009, the phase of variability was stable while the average magnitude and the amplitude changed (Fig.~\ref{asas}). For example, in 2002--2004 the amplitude was minimal while later began to increase following the activity. There were time spans when the variability is disappeared almost during one period at least. In bottom of Fig.~\ref{asas} shown the phase curve for data sets in 2005--2009 when the amplitude was more stable. The main trend was subtracted. A dispersion of data ($\sigma_{m_V }\approx 0.02^m$) is better than announced by authors for individual measurements. It is caused by photometry errors for an axis of ordinate and, possible also, by a changing of the active longitude within $\sim$0.1 of the period for an axis of abscissa. It corresponds to $\sim$36$\deg$ on the PZ~Mon surface. However, we cannot separate these two effects due to a similar contribution. This is possible in case of accurate photometry. 

The new observations were taken in the $BVR_C$-bands with a CCD imaging system installed at the 1.25-m telescope AZT-11 of the Crimean Astrophysical Observatory during 23 Jan -- 5 Apr 2015. In this interval 21 dates of observations were received, in each date it was carried out a few records, one record includes a sequence of objects image registration in each of three filters with a time resolution of about 3 minutes. The full dataset contains 94 measurements of brightness,  which cover about two periods of PZ~Mon. The obtained dense set of data is distributed on the most of phases of one period. Today it is the first photometric observations of PZ~Mon that have given an obvious evidence of its existence with a well accuracy. All magnitudes were reduced to the standard system and for the each date the $BVR$ values were averaged and the mean time of records was determined. Accuracy of observations had estimated from a data set of the comparison star HD~49477, it does not exceeds of 0.007$^m$ in the $V$-band. In this paper we analyse data of the $V$-band only.

\articlefigure{2015.eps}{cmp}{Photometry of PZ~Mon in 2015 (\textit{circles}). \textit{Solid line} is ephemeris data limited by error of the early estimated period (\textit{grey line}).}

Fig.~\ref{cmp} represents the photometric data in the $V$-band overlapped by ephemeris calculated by \citet{2015AstL...41..677P} in assumption of the big symmetrical spotted area. In this case, we can use cosines law with the average magnitude and the amplitude corresponding to observable values: $m_V = 9.17+0.05\,\textrm{cos}(2\pi\,JD/P)$, where $JD=2454807.2+34.13\,E$. The shift between them is obvious which may be explained by two ways. The first way is a correction of the period. Indeed, scattered data of the ASAS photometry does not allows to reach for better accuracy, and the corrected value of 34.12~days, which in limits of error, describes the observations in 2015 more accurate. However, there is a second way. The found shift is about of 3~days corresponds to 0.09 of the period or $\sim$32$\deg$ on the PZ~Mon surface. The same value was estimated using ASAS data. Note, early we found the size of a spotted area close to the size of hemisphere \citep{2015AstL...41..677P}. So, value 32$\deg$ can be explain by differential rotation or spot distribution on the spotted hemisphere without a correction of the period.

A form of light curve is slightly different from cosines law constructed using ephemeris in Fig.~\ref{cmp}. It specifies on a small asymmetry of  shape of the spotted area along longitudes. Nevertheless, the location of the activity region has no significant changes, always being on the side located under the secondary component. This observable fact is derived using new accurate photometry confirms our conclusion that activity of PZ~Mon is caused by the small component. The same source of the activity is observable for several RS~CVn stars with the secondary component of small mass \citep{2015AstL...41..677P}. However, they are asynchronous systems while PZ~Mon is synchronous, so a question of the nature of synchronisation is remain.

\acknowledgements This investigation was supported by Basic Research Program P-7 of the Presidium of the Russian Academy of Sciences.


\begin{thebibliography}{}
\bibitem[Pakhomov et al., (2015)]{2015MNRAS.446...56P} Pakhomov, Yu.~V., Chugai, N.~N., Bondar', N.~I., Gorynya, N.~A. \& Semenko, E.~A. 2015, MNRAS, 446, 56
\bibitem[Pakhomov, (2015)]{2015AstL...41..677P} Pakhomov, Yu.~V 2015, Astronomy Letters, 41, 677
\bibitem[Bondar', (1995)]{1995AAS..111..259B} Bondar', N.~I. 1995, A\&AS, 111, 259
\bibitem[Bondar' and Prokof'eva, (2007)]{2007OAP....20...14B} Bondar', N.~I. \& Prokof'eva, V.~V. 2007, Odessa Astronomical Publications, 20, 14 
\bibitem[Pojmanski, (1997)]{1997AcA....47..467P} Pojmanski, I. 1997, Acta Astron., 47, 467
\end{thebibliography}


\end{document}